\documentclass[12pt]{article}
\usepackage{amssymb,amsmath,bm,here,amssymb,epsfig,here}

\usepackage{amssymb,amsmath,amsfonts,amsbsy,graphicx}
\usepackage{color}
\usepackage{psfrag}
\renewcommand{\thefootnote}{\fnsymbol{footnote}}

\newcommand{\re}{{\rm Re\,}}

\newcommand\ba{\begin{eqnarray*}}
\newcommand\ea{\end{eqnarray*}}

\newif\iffigure
\figuretrue

\begin{document}
\title{}

\title{
\begin{flushright}
\begin{minipage}{0.2\linewidth}
\normalsize
CTPU-17-23\\
EPHOU-17-010 \\
WU-HEP-17-12 \\*[50pt]
\end{minipage}
\end{flushright}
{\Large \bf 
A viable D-term hybrid inflation\\*[20pt] } }

\author{Kenji~Kadota$^{1,}$\footnote{
E-mail address: kadota@ibs.re.kr}, \ 
Tatsuo~Kobayashi$^{2,}$\footnote{
E-mail address: kobayashi@particle.sci.hokudai.ac.jp} \ and \ 
Keigo~Sumita$^{3,}$\footnote{
E-mail address: k.sumita@aoni.waseda.jp
}\\*[20pt]
$^1${\it \normalsize 
Center for Theoretical Physics of the Universe,
Institute for Basic Science (IBS),}\\
{\it \normalsize Daejeon 34051, Korea} \\
$^2${\it \normalsize 
Department of Physics, Hokkaido University, 
Sapporo 060-0810, Japan} \\
$^3${\it \normalsize 
Department of Physics, Waseda University, 
Tokyo 169-8555, Japan} \\*[50pt]}

\date{
\centerline{\small \bf Abstract}
\begin{minipage}{0.9\linewidth}
\medskip 
\medskip 
\small 
We propose a new model of the D-term hybrid inflation in the framework 
of supergravity.
Although our model introduces, analogously to the conventional D-term inflation, the inflaton and a pair of scalar fields charged under a $U(1)$ gauge symmetry, we study the logarithmic and exponential dependence on the inflaton field, respectively, for the K\"ahler and superpotential.
This results in a characteristic one-loop scalar potential consisting of linear and exponential terms, which realizes the small-field inflation dominated 
by the Fayet-Iliopoulos term. 
With the reasonable values  for the coupling coefficients and, in particular, 
with the $U(1)$ gauge coupling constant comparable 
to that of the Standard Model, our D-term inflation model can solve the notorious problems in the conventional D-term inflation, namely, 
the CMB constraints on the spectral index and the generation of cosmic strings.
\end{minipage}}

\begin{titlepage}
\maketitle
\thispagestyle{empty}
\clearpage
\tableofcontents
\thispagestyle{empty}
\end{titlepage}

\renewcommand{\thefootnote}{\arabic{footnote}}
\setcounter{footnote}{0}

\section{Introduction}
The slow-roll inflation is a successful paradigm to describe our universe 
and various models have been proposed so far, along with 
actively updated data of Cosmic Microwave Background (CMB) observations. 
In the context of supergravity,  inflation models can be classified into two classes depending on whether the inflation energy scale is dominated by the F-term or D-term. 
An advantage of F-term models is that a wide variety of the forms of the inflaton potential and consequently the inflation dynamics are possible by varying the K\"ahler potential and superpotential. 
The F-term models, however, generically suffer from a notorious problem, 
the so called $\eta$ problem. The central issue in the model building based on the F-term is 
how to protect the inflaton potential against supergravity corrections which give the inflaton 
a heavy mass at least of order the Hubble scale. 
On the other hand, the inflation models based on the D-term potential 
are not exposed to the $\eta$ problem spoiling the flatness of the potential  \cite{Halyo:1996pp,Binetruy:1996xj}. 
The D-term inflation models are based on 
a $U(1)$ gauge theory with a Fayet-Iliopoulos (FI) term, 
and its D-term provides the dominant inflationary energy density. 
Recent CMB observations, however, contradict the predictions of the original D-term hybrid inflation model and, in particular, there exit persistent conflicts for the spectral tilt and the cosmic strings. 
The CMB spectral tilt for the conventional D-term inflation is predicted as
\begin{equation*}
n_s\sim1-\frac1N. 
\end{equation*}
Substituting typical values of the efolding number $N = 50-60$, 
we find that the predicted value of $n_s$ lies outside of the $2\sigma$ range 
of the recent observations\cite{Ade:2015xua,Ade:2015lrj}. 
The other problem is caused by the local cosmic strings 
which are formed at the end of 
the inflation when the $U(1)$ gauge symmetry is broken \cite{Lyth:1997pf,Jeannerot:1997is}. 
The cosmic strings contribute to the CMB anisotropies\cite{Gott:1984ef,Kaiser:1984iv}, 
and a stringent restriction appears. The D-term hybrid model generically 
predicts the magnitude of the contributions much larger than 
the observational upper limit.

There have been many attempts to modify the inflationary dynamics 
in order to revive the D-term hybrid model by considering more complicated potentials or 
additional dynamics\cite{Rocher:2006nh,Lin:2006xta,Seto:2005qg,Endo:2003fr,
Domcke:2014zqa,Evans:2017bjs}. 
Nevertheless it is not straightforward to simultaneously obtain the desirable spectral index and a sufficiently small cosmic string tension to rescue the D-term inflation model, 
and, in most cases, extremely small values of coupling constants 
including the gauge coupling are required. 
In view of an ever growing amount of cosmological data, the D-term inflation would be now on the verge of exclusion.


In this paper, we propose a new model for the D-term hybrid inflation, 
where all the observational constraints are satisfied by 
a reasonable choice of the parameter values. 
This paper is organized as follows. After reviewing the conventional D-term hybrid 
inflation model in Section 2, we present the essential features of our new model in Section 3 where we demonstrate that the model is capable of realizing 
the successful inflation along with concrete examples for the illustration. 
Section 4 investigates the sensitivity of the inflation dynamics on the model parameters.
We will comment on the scalar potential after the inflation 
and its vacuum structure in Section 5, 
and Section 6 is devoted to the conclusion 
and discussion.

\section{Conventional D-term inflation}

We first briefly review the conventional D-term inflation model for comparison to clarify the notable features of our D-term inflation model.
The conventional D-term inflation includes three fields, $X$ and $\phi_{\pm}$.
The K\"ahler potential is written by
\begin{equation*}
K = |X|^2 + |\phi_+|^2 + |\phi_-|^2,
\end{equation*}
and one uses the following superpotential:
\ba
W=\lambda X \phi_+ \phi_- .
\ea
We use the convention that the superfield and its scalar component are written by the same letter.

We introduce a $U(1)$ gauge symmetry, and $\phi_{\pm}$ have the $U(1)$ charges, $\pm 1$, while 
$X$ has a vanishing $U(1)$ charge. 
Then, the $U(1)$ D-term is given by 
\ba
D=g\left(\xi+|\phi_+|^2-|\phi_-|^2\right), 
\ea
where $g$ and $\xi$ are the gauge coupling and the FI term.
Such a FI term could be obtained by moduli fields after moduli fields are stabilized 
by a certain mechanism, or generated  by some strong dynamics.
We assume that $\xi >0$.

The masses of $\phi_\pm$ are given by, for $X=\chi e^{i \theta}/\sqrt2$, 
\begin{equation*}
m^2_\pm=\lambda^2\chi^2/2 \pm g^2\xi .
\end{equation*}
For a certain value of $\chi$,  both $m^2_\pm$ can be positive and 
the charged fields are stabilized as $\phi_\pm=0$. 
The D-term, $D=g\xi$, then serves as the energy source of the inflation 
as long as $m^2_\pm>0$. 
The supersymmetry breaking effect produces the potential for the inflaton $\chi$ 
field through the quantum corrections. 
The inflation is viable 
when the loop-corrected potential allows the slow-rolling of the inflaton with $m^2_\pm>0$.
The field $\phi_-$ is destabilized for $\chi < \chi_{\rm end}=g \sqrt{2\xi}/\lambda$ signaling the end of inflation.
That is, $\phi_-$ is the so called waterfall field. 

The one-loop contributions of the mass splitting to the scalar potential are 
calculated by using the Coleman-Weinberg (CW) formula \cite{Coleman:1973jx}, 
\begin{equation}
V_{\rm CW}=\frac1{64\pi^2}\sum_i \left(-1\right)^{F_i}M_i^4\log
\left(\frac{M_i^2}{\Lambda^2}\right). \label{eq:cwformula}
\end{equation}
$\Lambda$ and $M_i$ denote the renormalization scale and the masses of component fields labeled by $i$ 
which runs over both of the bosons and the fermions. 
Symbol $\left(-1\right)^{F_i}$ takes $+1$ and $-1$ for the bosonic and fermionic contributions, respectively. 
For $\chi \gg \chi_{\rm end}$ in the conventional D-term inflation model, 
the CW potential at the leading order in $(\chi_{\rm end}/\chi)^2$ 
leads to the inflaton potential
\begin{equation}
V=\frac{g^2 \xi^2}{2}+\frac{g^4 \xi^2}{16\pi^2} \ln \left( \frac{\lambda \chi}{\Lambda}\right)^2 .
\label{eq:conpot}
\end{equation}
The F-term potential vanishes during the inflation because of 
$\phi_\pm=0$. 
The slow-roll parameters then read 
\ba
\epsilon \equiv \frac12\left(\frac{V'}{V}\right)^2 \approx \frac{g^4}{32 \pi^4 \chi^2}, \qquad \eta\equiv \frac{V''}{V}\approx -\frac{g^2}{4\pi^2 \chi^2}.
\ea
The ratio $|\epsilon/\eta|$ is suppressed by a loop factor as $|\epsilon/\eta| \approx g^2/8 \pi^2$.
We can estimate the CMB spectral tilt 
\ba
n_s-1 =-6\epsilon+2\eta\approx 2 \eta\approx  -1/N ,
\ea
where the last equality holds if the cosmologically relevant scales leave the horizon when $\chi \gg \chi_{\rm end}$ so that
\ba
N=\int^{\chi}_{\chi_{\rm end}}\frac{V}{V'}d\chi =\frac{2 \pi^2}{g^2}\left( \chi^2 -\chi_{\rm end}^2\right) \sim \frac{2 \pi^2}{g^2} \chi^2 .
\ea
Then, putting typical values of $N=50-60$, we obtain $n_s \gtrsim 0.98$ contradicting with the CMB data which prefers a smaller $n_s$.

The curvature perturbation amplitude is given by 
\ba
A_s=\frac{V^3}{12\pi^2(V')^2} = \frac{V}{24\pi^2 \epsilon} \approx \frac{\xi^2}{3(1-n_s)} .
\ea
Using $A_s$
we can see that the conventional D-term inflation leads to too big a value of string tension 
\begin{equation}
G \mu=2\pi\langle\phi_-\rangle^2=2\pi \xi\sim 4\times 10^{-6} \times
\left(
\frac{1-n_s}{0.03}
\right)^{1/2}
\left(
\frac{A_s}{2\times 10^{-9}}
\right)^{1/2} .
\label{gmufixed}
\end{equation}
The latest Planck observation found the upper limit of the string tension as 
$G\mu<3.3\times 10^{-7}$\cite{Ade:2015xua,Ade:2015lrj}. 
Thus, the above predicted value is too large.
Note that, in the conventional D-term inflation model, $\epsilon$ and $\eta$ are tightly correlated and their ratio is fixed for a given gauge coupling constant $|\epsilon/\eta|\approx  g^2/8 \pi^2$, so that
the string tension is fixed once the CMB power spectrum is given as shown in Eq. (\ref{gmufixed}).
In the next section, we will propose a new D-term inflation model to resolve these problems.

\section{New model for D-term inflation}
We now present a new model for the D-term hybrid inflation.
We study the following K\"ahler potential and superpotential, 
\begin{eqnarray*}
K&=&-x\log\left(T+\bar T\right)
+\frac1{\left(T+\bar T\right)^y}\left(|\phi_+|^2+|\phi_-|^2\right),\\
W&=&\lambda e^{-a T}\phi_+\phi_-\label{eq:kw1}, 
\end{eqnarray*}
where $x,y,a$ and $\lambda$ are model parameters. 
Note that we adopt the unit of the reduced Planck mass scale, 
$M_{\rm P} = 2.4\times10^{18}$ GeV.
The chiral superfield $T$ will be identified as an inflaton field.
We then consider a $U(1)$ gauge symmetry 
with a non-vanishing FI term 
under which $T$ and $\phi_\pm$ have charges of $0$ and $\pm1$, 
respectively. 
We canonically normalize charged fields $\phi_\pm$ and 
denote them by $\tilde\phi_\pm$. 
The D-term is then given by 
\begin{equation*}
D=g\left(\xi+|\tilde\phi_+|^2-|\tilde\phi_-|^2\right) .
\end{equation*}

Although the forms of our K\"ahler and superpotential resemble those obtained within the framework of superstring theory, we present our analysis treating our new model as an effective supergravity model deferring the UV completion for the future work (we give a brief comment on the realization of our model in superstring theory in the discussion section). For instance, similarly to the conventional D-term inflation, our FI term could be induced by a modulus field, which is 
  independent of $T$, or dynamically by strong interactions.\footnote{Another string theory construction would be also possible where 
  the modulus field plays a role of the inflaton and the FI term also depends on the same modulus \cite{Kobayashi:2003rx}.}




We assume the constant FI term and $\xi>0$.
The masses of $\tilde\phi_\pm$ are given by 
\begin{equation*}
m^2_\pm=\lambda^2\left(T+\bar T\right)^{-x+2y}e^{-a\left(T+\bar T\right)}
\pm g^2\xi. 
\end{equation*}
For a certain value of $T+\bar T$, $m^2_\pm$ can be positive and 
the charged fields are stabilized as $\phi_\pm=0$. 
Then, similarly to the conventional D-term hybrid inflation, 
the inflation can be realized as long as $m^2_\pm>0$. 
The CW potential (\ref{eq:cwformula}) with the same approximation as 
that used in Eq.~(\ref{eq:conpot}) provides 
the inflaton potential of the form\footnote{
The successful inflation will be eventually realized 
in a small field regime in our scenario, 
where it is non-trivial that we can approximate the CW potential 
by the present form. 
We have examined the complete form of the CW potential (\ref{eq:cwformula}) to confirm that 
the approximate potential presented in our discussions is satisfactory to reveal 
the essential features of the inflationary dynamics predicted by our model. 
Therefore, we will study the inflationary potential of this approximated form and the additional terms in the exact potential do not quantitatively affect our discussions within the parameter range of our interest.}, 
\begin{equation*}
V_{\rm inf}=\frac{g^2\xi^2}2+\frac{g^4\xi^2}{16\pi^2} \log\left[
\left(T+\bar T\right)^{-x+2y}e^{-a\left(T+\bar T\right)}\right],\label{eq:vinf}
\end{equation*}
where we have omitted the constant term such as $\log(\lambda/\Lambda)$.
We define the canonically normalized real scalar field $\tilde\tau$ 
by using a K\"ahler metric $K_{T\bar T}$, as 
\begin{equation*}
\frac{d\tilde\tau}{d\tau} =\sqrt{K_{T\bar T}},\qquad T=\frac{\tau+i\sigma}{\sqrt2}, 
\end{equation*}
that is, $T+\bar T=\sqrt2\exp\,(\sqrt2\tilde\tau/\sqrt x)$. 
In this paper, $\tilde\tau_{\rm CMB}$ and $\tilde\tau_{\rm end}$ are used to 
represent the field values at which the CMB fluctuation is created and 
the inflation ends, respectively. 
By using the normalized fields, $m^2_\pm$ and $V_{\rm inf}$ are expressed as 
\begin{equation}
m^2_\pm = \lambda^22^{\frac{-x+2y}2}\exp\left[-\sqrt2ae^{\sqrt\frac2x\tilde\tau}
+\sqrt\frac2x\left(-x+2y\right)\tilde\tau\right]\pm g^2\xi,
\label{eq:msquared}
\end{equation}
and 
\begin{equation}
V_{\rm inf} = \frac{g^2\xi^2}2+\frac{g^4\xi^2}{16\pi^2} \left[
-\sqrt2ae^{\sqrt\frac2x\tilde\tau}+\sqrt\frac2x(-x+2y)\tilde\tau\right]. \label{eq:infpotential}
\end{equation}

The slope of the inflaton potential is dominated by the linear term 
for $\tilde\tau\lesssim 0$, 
where the slope is given by its coefficient proportional to $-x+2y$. 
For the other region, $\tilde\tau\gtrsim0$, 
the exponential term is dominant and its behavior 
drastically depends on the sign of $a$. 
Different signs of $a$ and $-x+2y$ would hence result in the inflation scenarios 
with totally different consequences.

The change of the sign from a positive to a negative value for the mass squared of $\tilde\phi_-$
can realize the hybrid inflation, and it is non-trivial whether this can happen in the present model
because our CW potential is quite different from that in the conventional D-term inflation. 
We depict $m_-^2$ and $V_{\rm inf}$ as functions of the inflaton 
in Fig.~\ref{fg:pot}, where we vary the values of $(x,y)$ and $a$ while fixing the other parameters as 
$g=1.0$, $\lambda=1.0$ and $\xi=1.0$.
Note that we have chosen these values just for the illustration. 
We will show their realistic values at the end of this section 
and in the following section. 
\iffigure
\begin{figure}[t]
  \centering
\includegraphics[width=6cm]{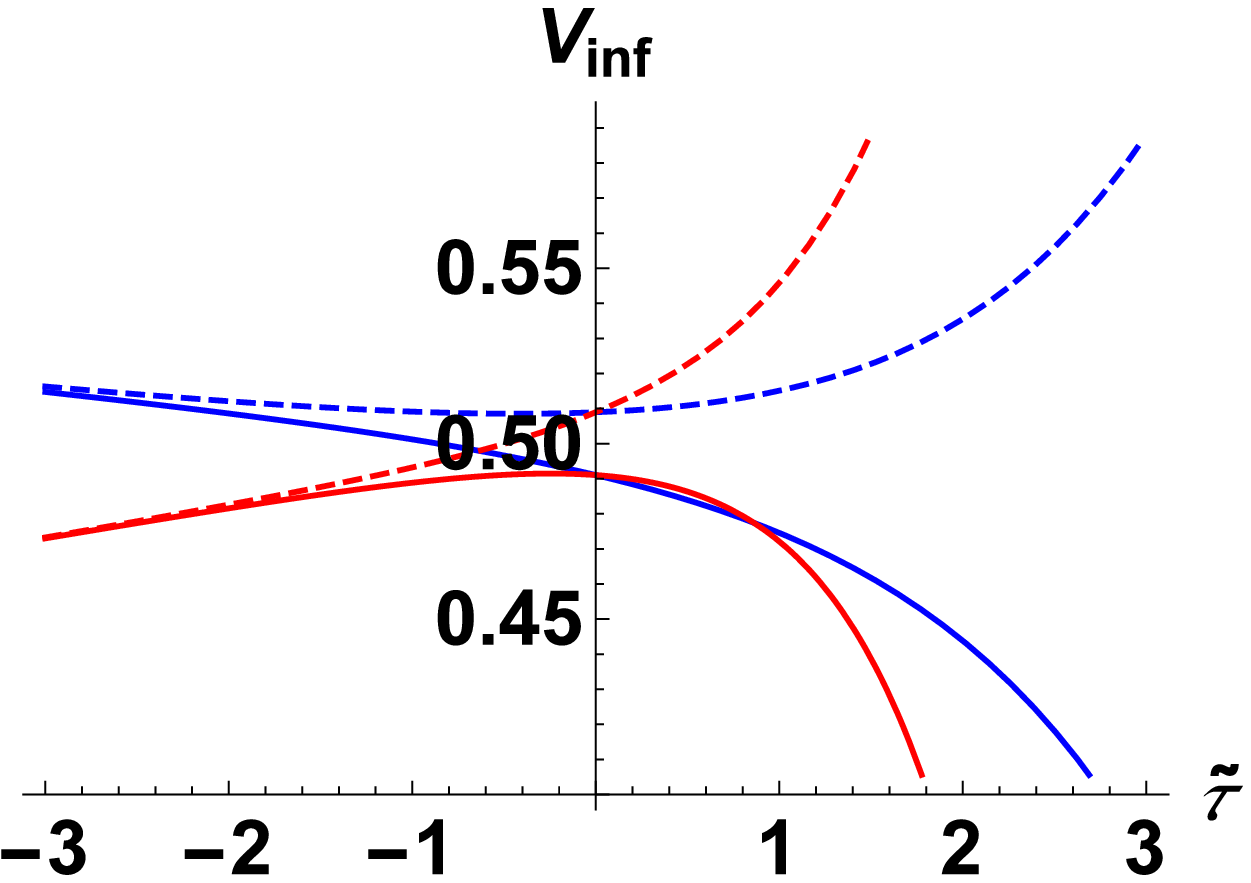}
\includegraphics[width=10cm]{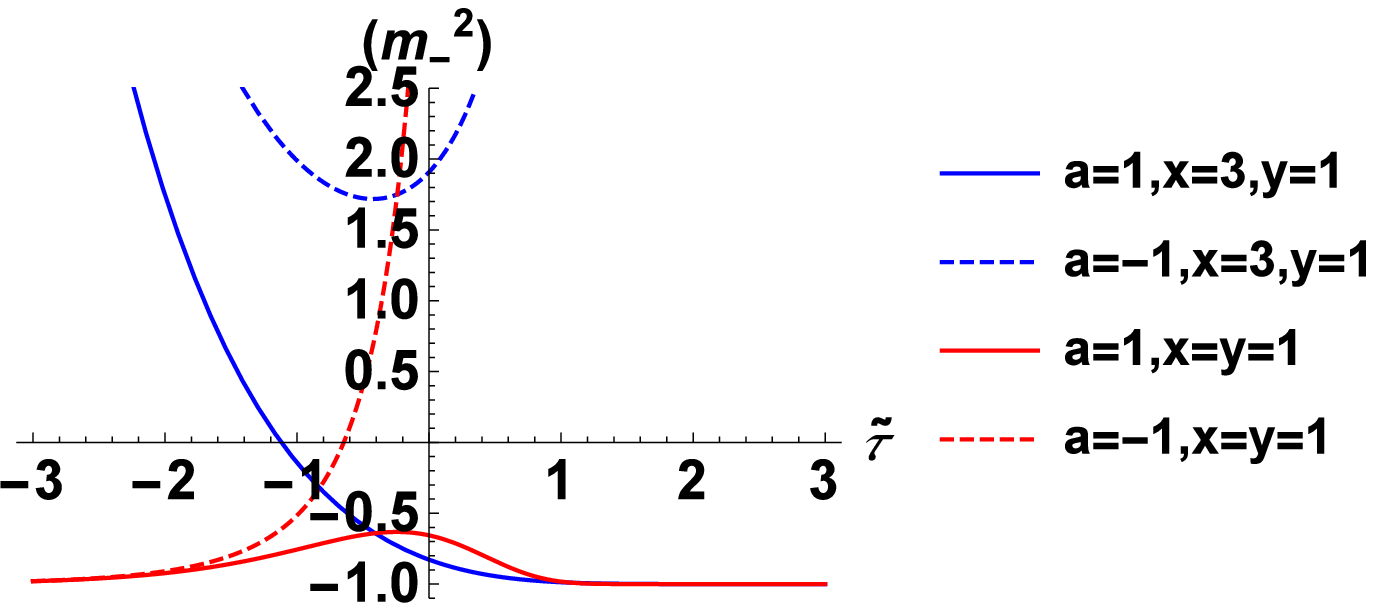} 
\caption{We show $m_-^2$ and $V_{\rm inf}$ 
with different values of $(x,y)$ and $a$, and the others are set as 
$g=1$, $\lambda=1$ and $\xi=1$. }\label{fg:pot}
\end{figure}
\fi 
Several possible inflaton trajectories are shown in this figure. For example, 
in the case with $a=1$, $x=3$ and $y=1$ (blue solid line), 
the inflaton can slowly roll down from $\tilde{\tau}<0$ to $\tilde{\tau}>0$
during which $m_-^2$ decreases. 
Note that only the decreasing behavior of $m_-^2$ 
is important here and we are not interested 
in the specific values of $m_-^2$ or the inflaton amplitude when the inflation ends
because those are freely controlled by $\lambda$ 
as seen from Eq.~(\ref{eq:msquared}) (also see the quantitative discussions in Section 4).

We calculate the slow-roll parameters in our model as 
\begin{eqnarray}
\epsilon&\equiv&\frac12\left(\frac{V'}{V}\right)^2=\frac12\left(\frac{g^2}{8\pi^2}\left(
-\frac{2a}{\sqrt x}\exp[\sqrt{\frac2x}\tilde\tau]+\sqrt{\frac2x}\left(-x+2y\right)\right)\right)^2,
\nonumber\\
\eta&\equiv&\frac{V''}{V}=-\frac{g^2a}{2\sqrt2\pi^2x}\exp[\sqrt{\frac2x}\tilde\tau]
\label{eq:eta}.
\end{eqnarray}
The spectral tilt and the tensor-to-scalar ratio are represented 
by these parameters as 
\begin{eqnarray*}
n_s&=&1-6\epsilon+2\eta,\\
r&=&16\epsilon. 
\end{eqnarray*}
The curvature perturbation amplitude is given by
\begin{equation*}
A_s=\frac{V^3}{12\pi^2(V')^2}=\xi^2\times\frac{8\pi^2}{3g^2}
\left(\sqrt{\frac2x}\left(-x+2y\right)-\frac{2a}{\sqrt x}
e^{\sqrt{\frac2x}\tilde\tau}\right)^{-2} .
\end{equation*}
By using the Planck normalization 
($A_s=2.2\times10^{-9}$), 
we find 
\begin{equation}
\xi=9.14\times10^{-6}\times g\left|\sqrt{\frac2x}\left(-x+2y\right)-\frac{2a}{\sqrt x}
e^{\sqrt{\frac2x}\tilde\tau}\right|. \label{eq:xi}
\end{equation}
The size of the cosmic string contributions to the CMB power spectrum 
is then estimated as 
\begin{equation}
G\mu =2\pi\langle\phi_-\rangle^2=2.29\times 10^{-6}\times g\left|\sqrt{\frac2x}\left(-x+2y\right)-\frac{2a}{\sqrt x}
e^{\sqrt{\frac2x}\tilde\tau}\right|\label{eq:gmu}, 
\end{equation}
where $\langle\phi_-\rangle^2=\xi$ after the inflation. 
This expression is given in the unit of the Planck mass scale 
because it is convenient to compare with the observational data. 

We first consider the case where the slope of the inflaton potential is 
dominated by the linear term. 
The cosmic string contribution is then approximated by 
\begin{equation*}
G\mu\sim2.29\times 10^{-6}\times g\left|\sqrt{\frac2x}\left(-x+2y\right)\right|, 
\end{equation*}
and the current CMB bound requires $g\lesssim0.1$ because $x$ and $y$ are of $\mathcal O(1)$ in general. 
In this case, the slow roll parameters are suppressed as 
\begin{eqnarray}
\epsilon&\lesssim&8.0\times10^{-9},\label{eq:epsbound}\\
|\eta|&\lesssim&1.3\times10^{-4}.\label{eq:etabound}
\end{eqnarray}
These cannot realize the observed value of $n_s$.

We next discuss the case with the dominant exponential term, that is, 
\begin{equation*}
\left|\frac{2a}{\sqrt x}
e^{\sqrt{\frac2x}\tilde\tau}
\right|>\left|\sqrt{\frac2x}\left(-x+2y\right)\right|
\sim\mathcal O(1). 
\end{equation*}
The contribution of cosmic strings is then given by 
\begin{equation*}
G\mu\sim2.29\times 10^{-6}\times g
\left|\frac{2a}{\sqrt x}
e^{\sqrt{\frac2x}\tilde\tau}\right|, 
\end{equation*}
and we find 
\begin{equation*}
g\times\left|\frac{2a}{\sqrt x}
e^{\sqrt{\frac2x}\tilde\tau}
\right|\lesssim0.1
~~\Leftrightarrow~~ g\lesssim0.1. 
\end{equation*}
By using this, we find the same upper limits 
on the slow-roll parameters as Eqs.~(\ref{eq:epsbound}) and 
(\ref{eq:etabound}) again, which cannot be consistent with the observed value of $n_s$.

The only way to realize the successful inflation is to consider 
a combination of the linear term and the exponential term, so that a cancellation between these two terms flattens the inflaton potential. 
Such a cancellation can happen only when the signs of $-x+2y$ and $a$ 
are the same. Otherwise, the slope of the potential would be enhanced. 
It is also important that the CW potential contains 
the linear term which contributes only to the slope of the potential and 
its second derivative vanishes. 
Only the exponential term contributes to the second derivative of the potential with respect to the inflaton 
while the slope is controlled by both terms. 
We can enhance the magnitude of $V''$ while keeping the flat slope. 
This is the crucial feature for the successful inflation in our model.

In generic hybrid inflation models, 
$\epsilon$ is trivially small enough to be neglected in the estimation of 
the spectral tilt. 
This is true also in the present model as demonstrated later. 
The value of $\eta$ hence should be negative 
in order to realize the observed value of $n_s$. 
As a result, we find $a>0$ because all the other factors appearing 
in Eq.~(\ref{eq:eta}), such as $g^2$ and $x$, are positive. 
Thus, the successful inflation is realized only when $-x+2y>0$ and $a>0$, 
which corresponds to the red solid lines in Fig.~\ref{fg:pot}. 
In this case, a hill-top like potential for the inflaton direction 
is obtained. The mass squared $m_-^2$ takes a maximum value at the top 
of the inflaton potential and it decreases as 
the inflaton is deviated from the peak. 
The hybrid inflation occurs when the inflaton is initially located around the top (we show the full potential with realistic values of the input 
parameters in Fig.~\ref{fg:pot4}).

A hill-top like potential for the inflaton 
is obtained for $-x+2y>0$ and $a>0$. 
We find that the top of this potential is represented by 
\begin{equation*}
\tilde\tau_0=\sqrt{\frac x2}\log\left[
\frac{-x+2y}{\sqrt2 a}
\right]\label{eq:tcmb}, 
\end{equation*}
which gives $V'=0$ with $\tilde\phi_\pm=0$. 
Around $\tilde\tau_0$, the potential is flattened to be 
suitable for the inflation. Fortunately, 
the inflaton amplitude at the end of inflation (where the inflation ends in a waterfall manner
due to $m_-^2$'s sign change from a positive to a negative value) can be freely controlled by $\lambda$ in our model without 
affecting the above discussions 
(the one-loop CW potential contains $\lambda$ but that contributes only 
to the height of the potential and that is much smaller than the contribution of 
the tree-level D-term potential). 
It is then assured that the inflation can always take place 
around this flattened top of the potential by choosing a proper value of $\lambda$. We will come back to the evaluation of the specific model parameters for the successful inflation later and we first present the following qualitative discussions
for the inflation which occurs around  $\tilde\tau_0$. 
We can estimate an approximate value of the spectral tilt by evaluating it at $\tilde{\tau}=\tilde\tau_0$  
\begin{equation*}
n_s\sim1+2\eta=1+\frac{g^2}{2\pi^2}\left(1-2\frac{y}{x}\right), 
\end{equation*}
which depends only on $g$ and $y/x$. 
We depict an allowed region on $(g,y/x)$-plane in Fig.~\ref{fg:gxy}. 
\iffigure
\begin{figure}[t]
  \centering
\includegraphics[width=9cm]{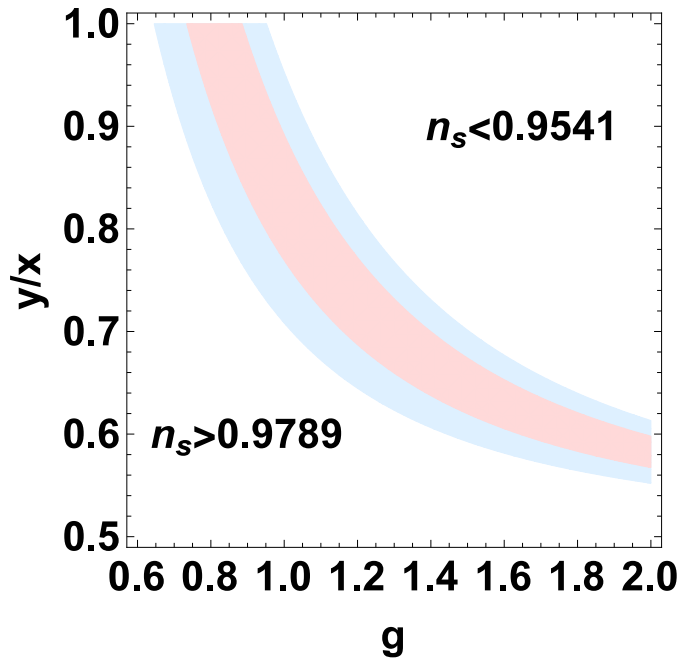}
  \caption{The dependence of $n_s$ on $g$ and $y/x$. 
In the pink (blue) region, 
the estimated value lies in the $1\sigma$ ($2\sigma$) 
range of the Planck observation. 
}\label{fg:gxy}
\end{figure} 
\fi
In the pink (blue) region, 
the CMB constraint on $n_s$ is satisfied 
at the $1\sigma$ ($2\sigma$)-level \footnote{Note that we assume $\tilde\tau_{\rm CMB}=\tilde{\tau_0}$ to draw this figure even though the actual $\tilde\tau_{\rm CMB}$ should be slightly deviated from $\tilde\tau_0$. This simplification suffices for our purpose of illustrating the sensitivity of $n_s$ on our model parameters. See the following discussions for more precise evaluation of our model parameters treating $\tilde\tau_{\rm CMB}\neq\tilde{\tau_0}$.}. 
The values of $g$ and $y/x$ in the colored region satisfy 
the observations. The most important thing seen from this figure is 
that the gauge coupling must be of $\mathcal O(1)$. 
This is in stark contrast to the conventional D-term inflation which 
forces the gauge coupling to be unnaturally small to satisfy the CMB constraints. 

Now let us go back to Eq. (\ref{eq:gmu}). 
The gauge coupling $g\sim\mathcal O(1)$ requires a slight cancellation between
the two terms in Eq. (\ref{eq:gmu}) to satisfy the CMB bound of $G\mu<3.3 \times10^{-7}$, 
which can be achieved when $\tilde\tau_{\rm CMB}\sim \tilde\tau_0$ because $G\mu$ is proportional to $V'$. 
Our model hence can vary $\lambda/\sqrt\xi$ and $\xi$, respectively, 
to obtain the desirable values of $\tilde\tau_{\rm end}$ 
and the inflation energy scale. This consequently indicates that our scenario can let, 
by adjusting our model parameters $\lambda$ and $\xi$, 
the cosmologically interesting scales leave the horizon at any desirable number 
of e-folds $N$ before the end of inflation, 
in a big contrast to the conventional D-term inflation model where the observed spectral index 
fixes $N$ which cannot be varied by adjusting the model parameters as reviewed in Section 2. Further quantitative discussions are given in Section 4.

We now illustrate the above qualitative discussions more quantitatively by the following concrete parameter set
\begin{equation}
(x,y)=(1,1),\qquad g=0.8,\qquad a=1.1,\qquad \lambda=1.4 \sqrt\xi\label{eq:in1}. 
\end{equation}
Fig.~\ref{fg:pot4} shows the normalized potential $V/\xi^2$. 
\iffigure
\begin{figure}[t]
\centering
\includegraphics[width=14cm]{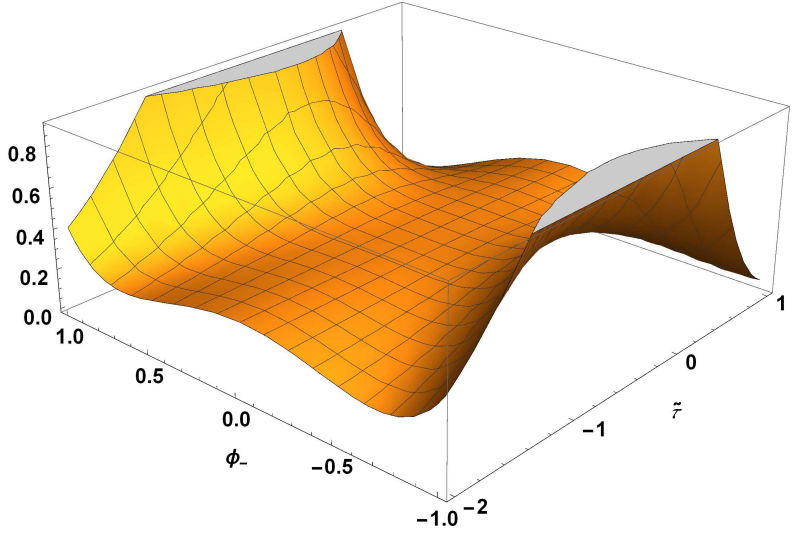}
\caption{We show $V/\xi^2$ with $(x,y)=(1,1)$, $a=1.1$, $g=0.8$, $\lambda=1.4\sqrt{\xi}$, 
which contains the tree-level F-term and D-term terms and 
the one-loop CW potential. 
}\label{fg:pot4}
\end{figure} 
\fi
The top of the potential is given by $\tilde\tau_0=-0.312$, 
where $\tilde\phi_-$ has a positive mass squared. 
It switches to a negative value as $\tilde\tau$ rolls away from 
$\tilde\tau_0$. 
We can find two transition points where $\tilde\phi_-$ becomes tachyonic 
around $\tilde\tau_0$.
That is, two different inflationary trajectories are available 
on this potential. 
First we focus on the trajectory where 
the inflaton rolls in the negative direction 
to arrive at the waterfall point given by $\tilde\tau_{\rm end}=-0.473$. 
In this case we find 
\begin{equation*}
\tilde\tau_{\rm CMB}=-0.388,\quad \sqrt \xi=1.03\times 10^{-3},
\quad n_s=0.971, 
\quad r=1.09\times 10^{-5},\quad G\mu=2.63\times 10^{-7}
\end{equation*}
for $N=50$, and 
\begin{equation*}
\tilde\tau_{\rm CMB}=-0.378,\quad \sqrt \xi=9.53\times 10^{-4},
\quad n_s=0.970, 
\quad r=8.12\times 10^{-6},\quad G\mu=2.27\times 10^{-7}
\end{equation*}
for $N=60$. 
In the other case, the inflaton rolls in the positive direction 
and arrive at $\tilde\tau_{\rm end}=-0.163$. 
We find 
\begin{equation*}
\tilde\tau_{\rm CMB}=-0.250,\quad \sqrt \xi=9.78\times 10^{-4},
\quad n_s=0.965, 
\quad r=8.98\times 10^{-6},\quad G\mu=2.39\times 10^{-7}
\end{equation*}
for $N=50$, and 
\begin{equation*}
\tilde\tau_{\rm CMB}=-0.260,\quad \sqrt \xi=8.95\times 10^{-4},
\quad n_s=0.965, 
\quad r=6.31\times 10^{-6},\quad G\mu=2.00\times 10^{-7}
\end{equation*}
for $N=60$. 
For reference we list the constraints on our model parameters from the Planck 
observation \cite{Ade:2015xua,Ade:2015lrj}, 
\begin{equation}
n_s=0.9665\pm0.0062,\quad r<0.11\quad{\rm and}\quad G\mu<3.3\times10^{-7}. 
\label{eq:obconst}
\end{equation}
It is worth emphasizing that the constrains on $n_s$ and $G\mu$ are satisfied by 
the reasonable values of input parameters (\ref{eq:in1}) in our model. 
The model realizes a similar consequence for any values of $g$ and $(x,y)$, as long as they lie in the $1 \sigma$ region of Fig.~\ref{fg:gxy}, because our freedom to tune the value of $\lambda$ without affecting the potential shape lets us use the flattened part of the potential for the desirable inflation dynamics.
We will investigate more generic features of our inflation model in the next section.

\section{Generic features}
In this section, we consider two sets of typical values of 
$(x,y)$ and $g$. One is given by $(x,y)=(1,1)$ and $g=0.8$, 
and the other is given by $(x,y)=(1,2/3)$ and $g=1.4$, 
both of 
which are located in the center of the $1 \sigma$ region shown 
in Fig.~\ref{fg:gxy}. 
We use Eq.~(\ref{eq:obconst}) as the observational constraints.

\subsection{Dependence on $\lambda$ } \label{sec:lambda}
It is important that, as discussed at the end of the last section, the desirable inflationary potential around $\tilde\tau_0$ 
is always available because the required value of $\tilde{\tau}_{\rm end}$ can be realized by a certain value of $\lambda$. 
We  investigate the specific value of $\lambda$ required for the purpose of illustrating this point. 
First, we convert the observational constraints 
on $(n_s,r)$, $A_s$ and $G\mu$ 
into the constraints on $\tilde\tau_{\rm CMB}$ and 
identify the corresponding allowed range of $\tilde\tau_{\rm CMB}$. 
Such an allowed range turn out to be $\tilde\tau_0\pm\mathcal O(0.1)$. 
More precisely, we varied $\tilde{\tau}_{\rm CMB}$ around $\tilde{\tau}_0$ for a given efolding number to find the allowed range of $\tilde\tau_{\rm end}$ as demonstrated in Fig.~\ref{fg:efo4050}. 
The different values of $(x,y)$ and $g$ are used 
in the left and right panels. 
\iffigure
\begin{figure}[t]
\centering
\includegraphics[width=8cm]{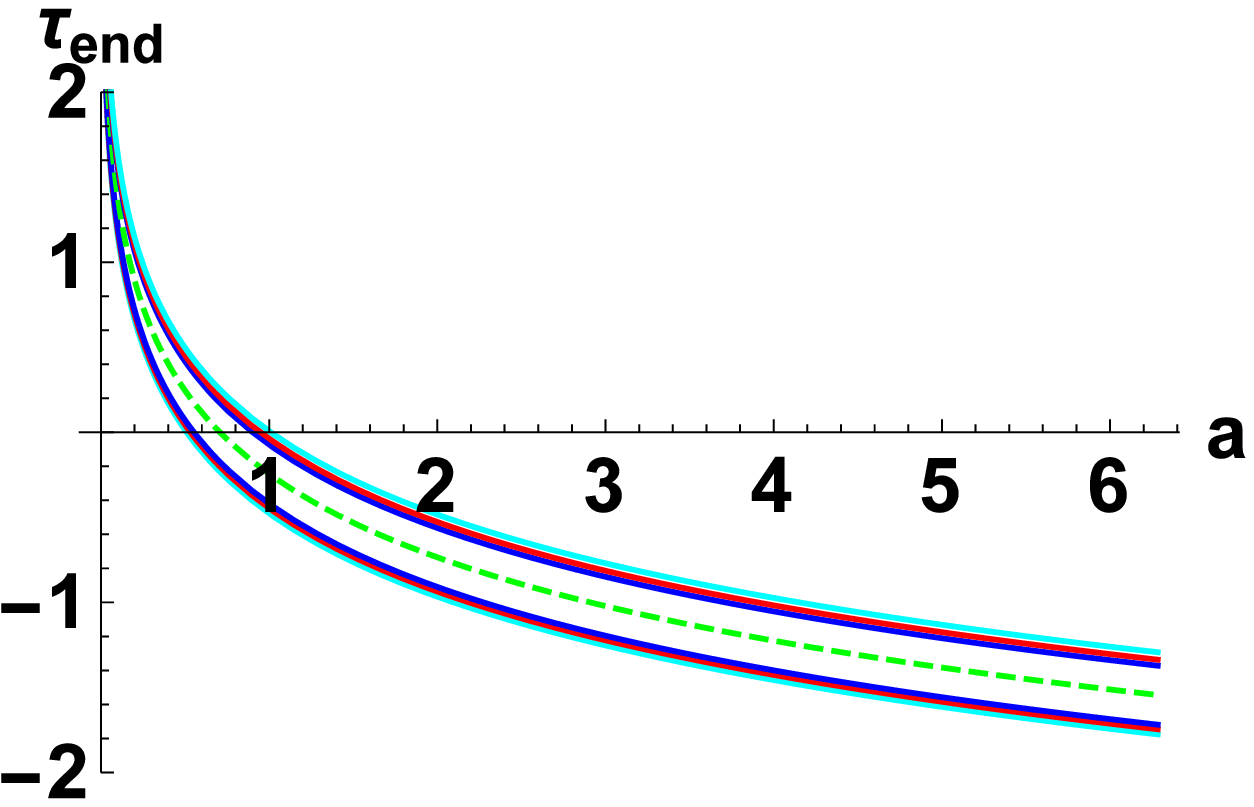}\hspace{0.5cm}
\includegraphics[width=8cm]{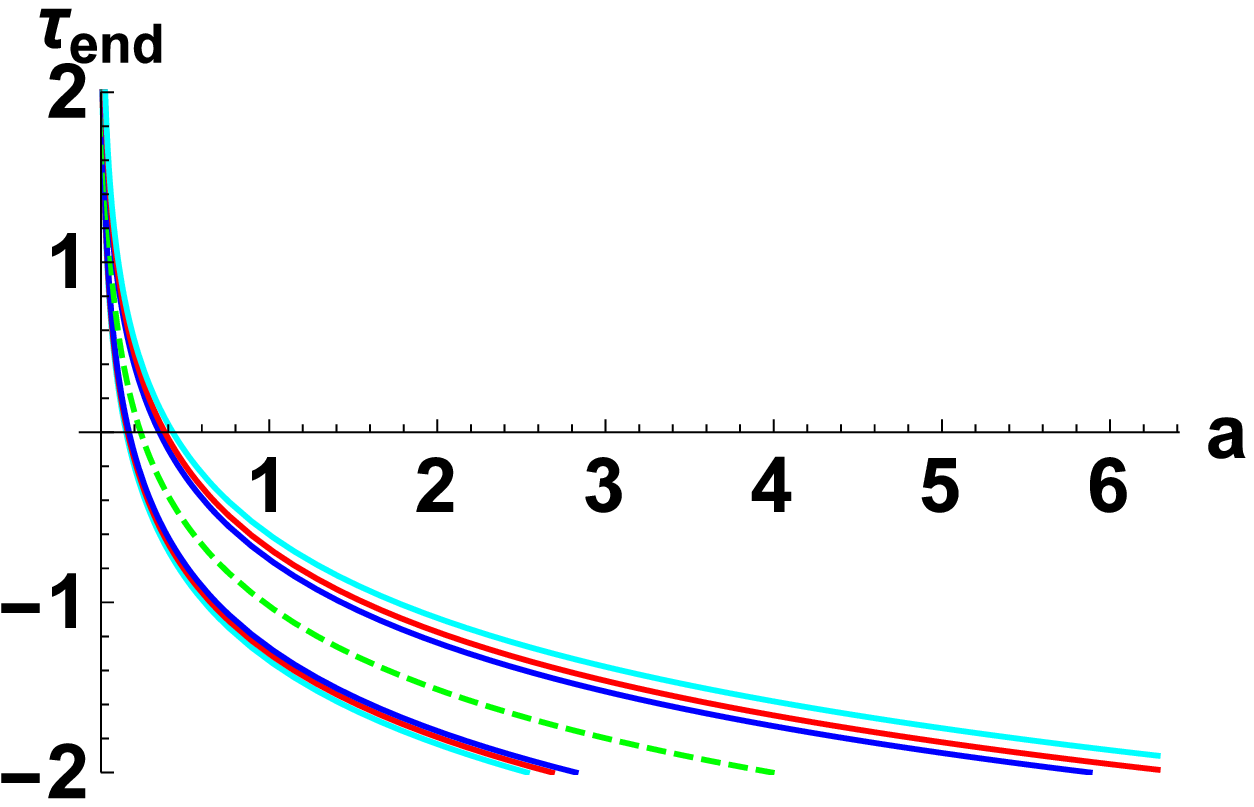}
\caption{We represent the observational upper and lower limits on $\tau_{\rm end}$ 
by the solid lines for $N=40$(Blue), $50$(Red) and $60$(Cyan). 
The parameters are set as $(x,y,g)=(1,\,1,\,0.8)$ and $(1,\,2/3,\,1.4)$ 
in the left and right panels, respectively. 
The green dashed line represents $\tilde\tau_0$. }\label{fg:efo4050}
\end{figure} 
\fi
The solid lines represent upper and lower limits 
on $\tilde\tau_{\rm end}$ for $N=40$(Blue), $50$(Red) and $60$(Cyan), 
and $\tilde\tau_0$ is indicated by the green dashed line. 
The present model always has two inflationary trajectories available, 
where the inflaton goes in the positive or negative direction. 
When $\tilde\tau_{\rm end}$ is located above (below) the green dashed line, 
the inflaton rolls in the positive (negative) direction.

We further explain the meaning of the solid lines. 
For example, $N=60$ is exactly required to satisfy the observational data 
on the cyan lines. In the inner region bounded by these lines, a smaller $N$ is allowed but $N=60$ still remains consistent with the CMB data unless 
the inflation energy scale becomes much smaller. Fig.~\ref{fg:efo4050} hence shows that there is no significant difference 
among these three values of the efolds, 
and in the rest of this subsection, we focus on the case with $N=60$.

Next we estimate the value of $\lambda$ required to realize 
the desirable value of $\tilde\tau_{\rm end}$ and it is shown in Fig.~\ref{fg:hoge1}. 
The value of the parameters are the same as before. 
The rainbow colored contours express the value of $\lambda/\sqrt\xi$. 
The gray shaded regions are excluded by the CMB observations. 
(The edge of the regions corresponds to the cyan lines 
in the last figure. )
The green dashed line represents $\tilde\tau_0$. 
The lower panels are magnification of a region where $\tilde\tau_{\rm end}>0$. 
We can conclude from these figures that 
the successful inflation can be realized by reasonable values of $\lambda$ with $\lambda/\sqrt{\xi}={\cal O}(1) $.

\iffigure
\begin{figure}[H]
\centering
\includegraphics[width=8cm]{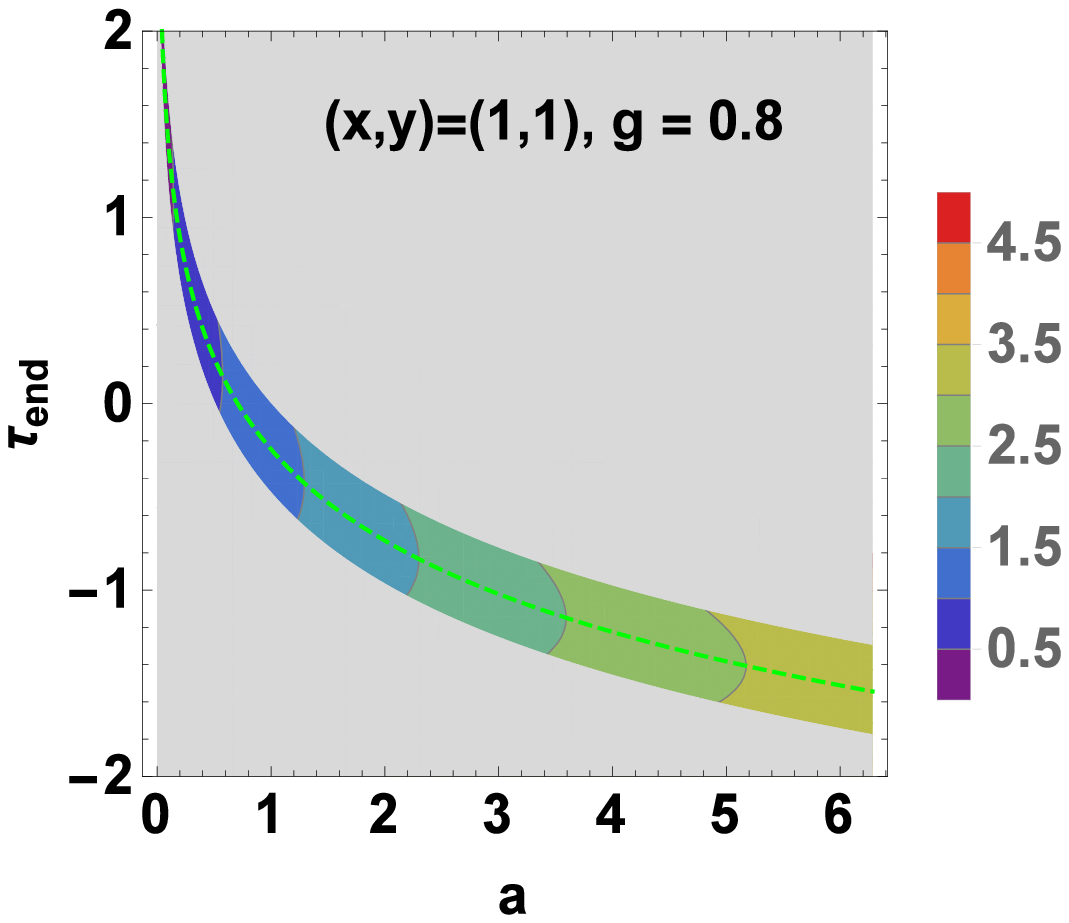}
\includegraphics[width=8cm]{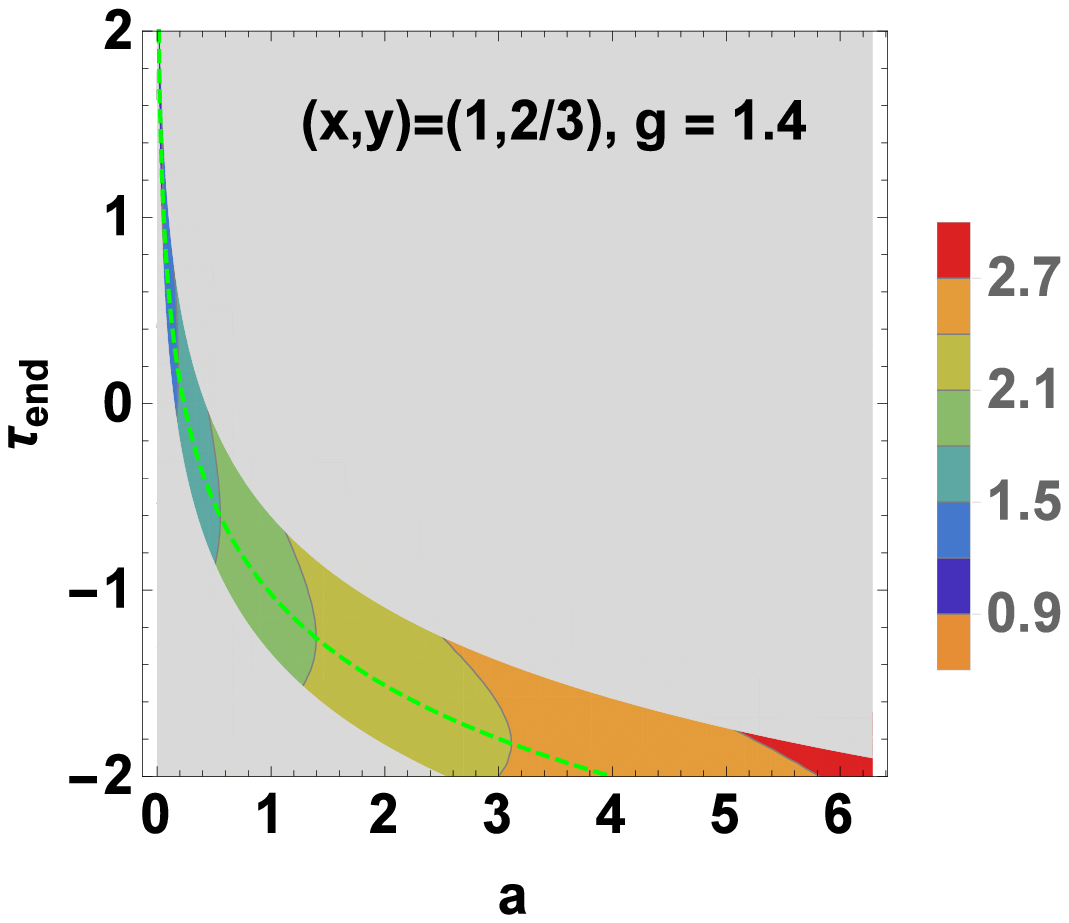}\\
\includegraphics[width=8cm]{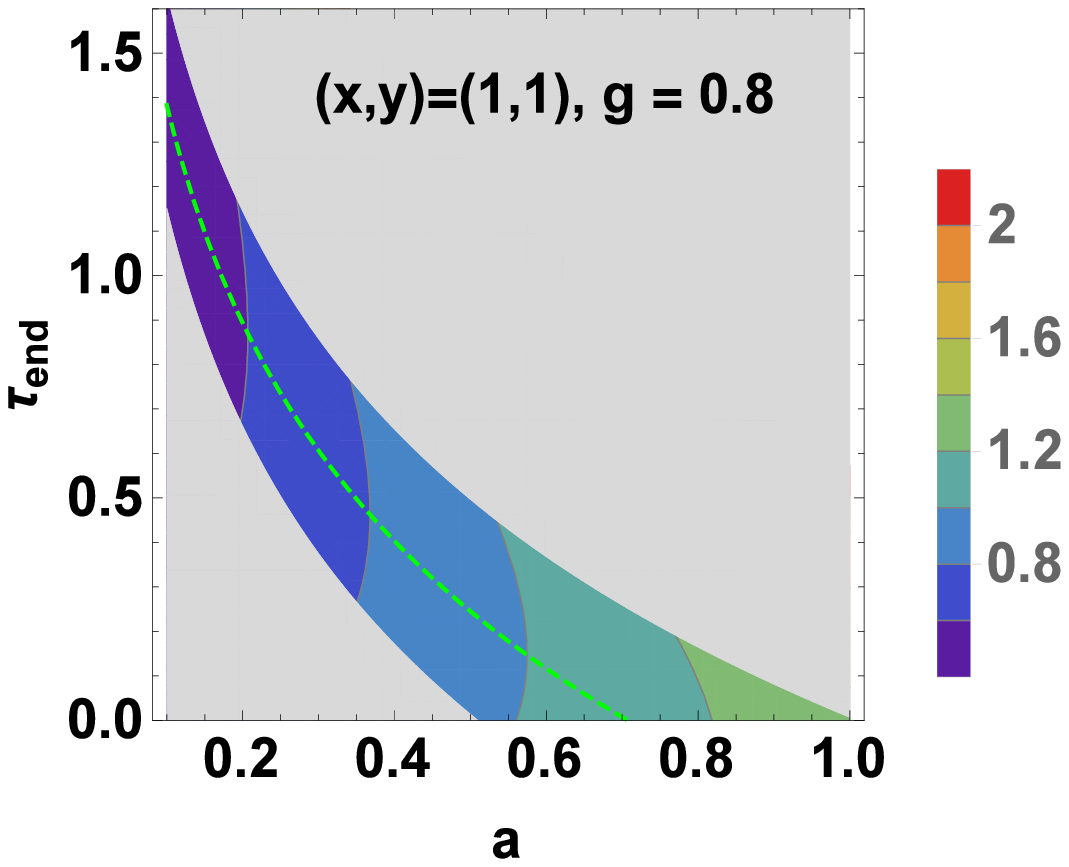}
\includegraphics[width=8cm]{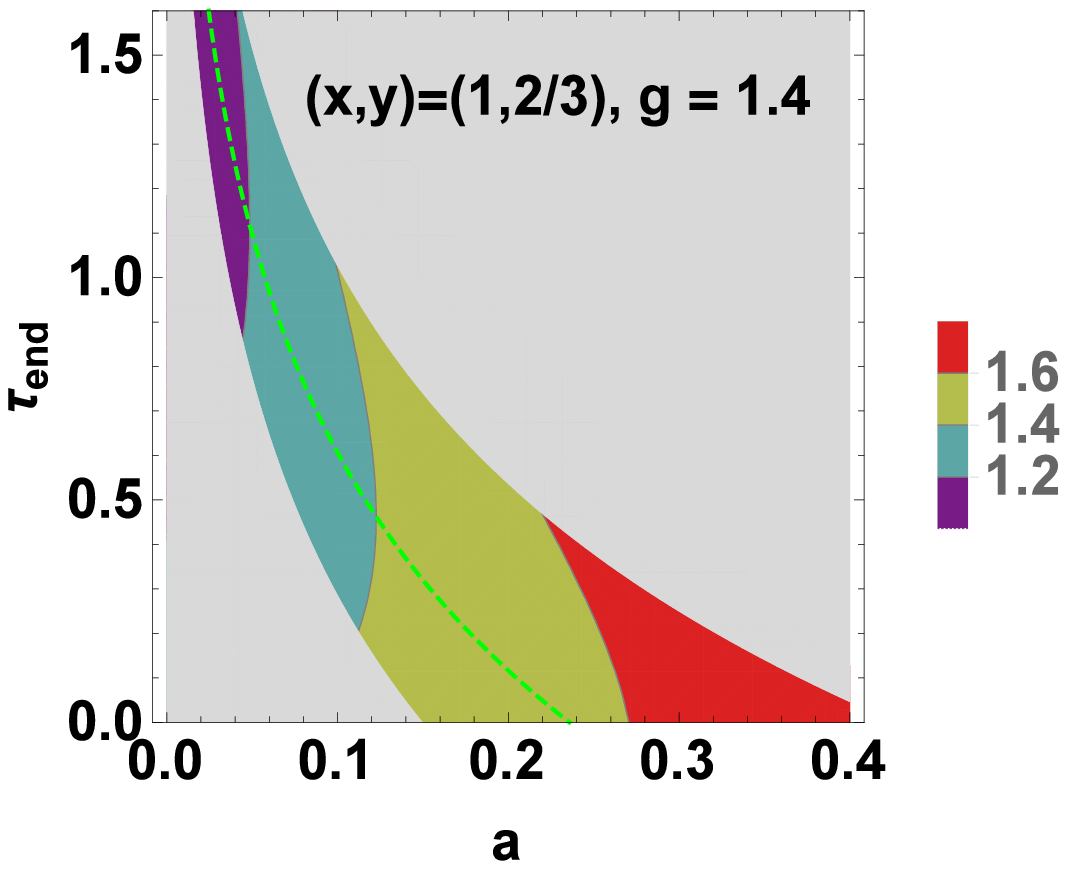}
\caption{We represent the values of $\lambda/\sqrt\xi$ by the contours 
in the upper panels with the different values of $(x,y)$ and $g$. 
The lower panels are the partial magnifications of the upper panels.
The gray shaded regions are excluded by the CMB observations. 
The green dashed line represents $\tilde\tau_0$ for $V'=0$. 
 }\label{fg:hoge1}
\end{figure} 
\fi

\subsection{Inflation scale: $H_{\rm inf}$}
We estimate the inflation energy scale, 
\begin{equation}
H_{\rm inf}\sim g\xi/\sqrt6 = 3.73\times 10^{-6} \times g^2 
\left|\sqrt{\frac2x}\left(-x+2y\right)-\frac{2a}{\sqrt x}
e^{\sqrt{\frac2x}\tilde\tau_{\rm CMB}}\right|,\label{eq:hinf} 
\end{equation}
where we have used Eq.(\ref{eq:xi}). 
In the last section, 
we found $g\sim \mathcal O(1)$ and 
the factor in the absolute value symbol of Eq.~(\ref{eq:hinf}) 
would be of $\mathcal O(0.1)$ 
to suppress the cosmic string contributions 
(see Eqs.~(\ref{eq:gmu}) and (\ref{eq:obconst})). 
Those lead to the typical scale of $H_{\rm inf}$ as 
$\mathcal O(10^{-6})$ or $\mathcal O(10^{-7})$. 

We investigate $H_{\rm inf}$ also in a numerical analysis, 
where we adopt the two ansatzs for $(x,y)$ and $g$ with $a=1$, 
and then, the remaining parameter in Eq.~(\ref{eq:hinf}) 
is only $\tilde\tau_{\rm CMB}$. 
We show its dependence on $H_{\rm inf}$ in Fig.~\ref{fg:loglog}. 
As mentioned in Section~\ref{sec:lambda}, 
the upper and lower bounds on $\tilde\tau_{\rm CMB}$ are given by 
$\tilde\tau_0\pm\mathcal O(0.1)$. 
Thus, we again find the typical scale of $H_{\rm inf}$ 
shown above, with $\tilde\tau_{\rm CMB}-\tilde\tau_0=0.1$. 
For $\tilde\tau_{\rm CMB}\sim\tilde\tau_0$, 
the inflaton potential is extremely flat and 
the inflation scale is drastically lowered (see, Eq.~(\ref{eq:xi})). 
One can see from this figure that a low scale inflation, e.g. 
$H_{\rm inf}\sim\mathcal O(1{\rm TeV})$, can happen 
in a small range of $\tilde\tau$ slightly apart from $\tilde\tau_0$ by 
$\mathcal O(10^{-10})$, which would require the fine-tuning of the initial condition.
In addition, note that a smaller  $H_{\rm inf}$ requires a smaller $\lambda \sim \sqrt{\xi} \sim \sqrt{H_{\rm inf}/M_{\rm P}}$, and we may need some suppression mechanism of $\lambda$ for the low scale inflation.
Although we here focused on a positive deviation from $\tilde\tau_0$, 
the analogous result follows for a negative deviation 
($\tilde\tau_{\rm CMB}-\tilde\tau_0<0$).

\iffigure
\begin{figure}[t]
\centering
\includegraphics[width=8cm]{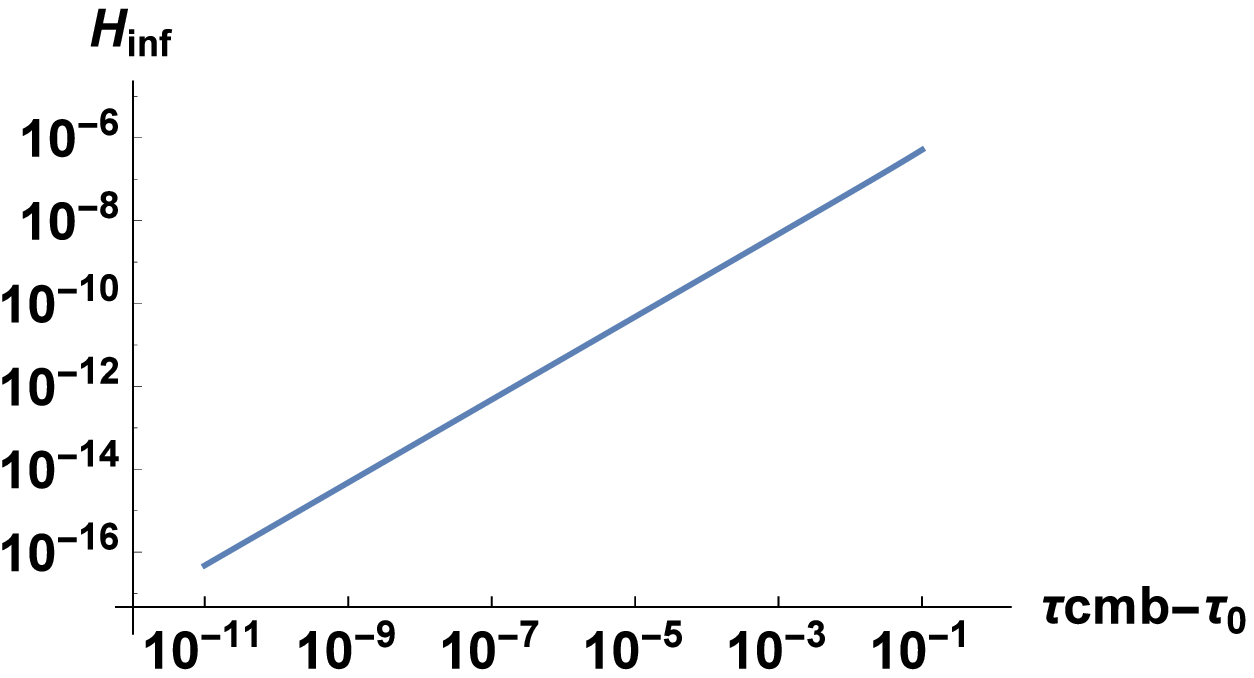}
\includegraphics[width=8cm]{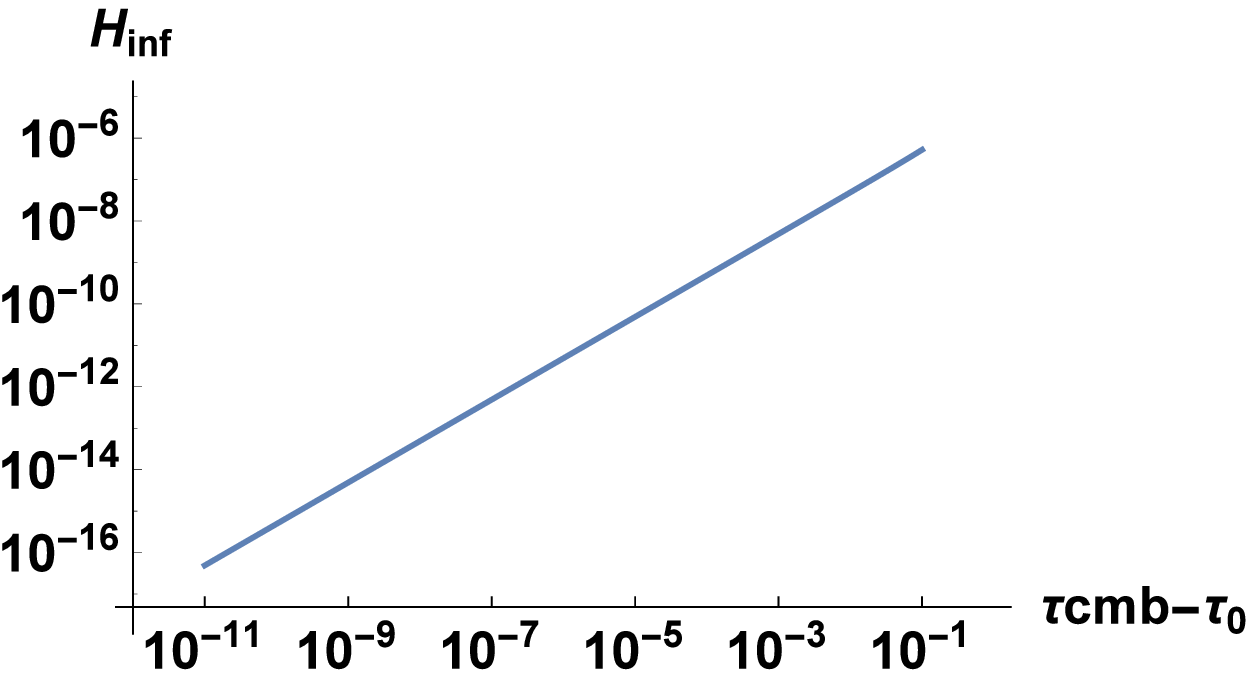}
\caption{We estimate the value of $H_{\rm inf}$ for $a=1$. The left and right panels correspond to 
$(x,y,g)=(1,1,0.8)$ and $(x,y,g)=(1,2/3,1.4)$, respectively.
}\label{fg:loglog}
\end{figure} 
\fi

\subsection{Dependence on $(x,y)$ } 
We discuss the impact of the value of $(x,y)$ on the inflation. 
As mentioned above, the inflaton potential we have investigated is realized by 
$-x+2y>0$ and we have assumed in our discussions $y/x>1/2$. 
If it is satisfied, the plausible inflationary potential 
can be realized by a proper value of $g$ 
as shown in Fig.~\ref{fg:gxy}. 
It is also important that $\tilde\tau_0$ depends on $x$ and $y$ rather than $y/x$. 
We may expect that the inflation happens in a range of positive field value of $\tilde\tau$ 
or $\re T >1$ when we embed the model into superstring frameworks. 
(Precisely, $\re T >1$ implies $\tilde\tau>0.245\sqrt{x}$.)
In that case the value of $a$ is constrained as $a<(-x+2y)/\sqrt2$. 
Since the value of $-x+2y$ can increase while fixing the ratio $y/x$, 
it is possible to realize the inflation with $\tilde\tau_{\rm end}>0$ 
for a wide range of $a$, although the flattened part of the potential 
or the allowed range of $\tilde\tau_{\rm end}$ 
would become small for a large value of $-x+2y$.

\section{Comments on the true vacuum} 
We remark on some issues after the inflation, where $\phi_-$ develops its vacuum expectation value and the F-term potential consequently arises.
For $\phi_+=0$ ($\xi>0$), we find 
\begin{equation*}
V_F=e^{-a(T+\bar T)+(T+\bar T)^{-y}|\phi_-|^2}\lambda^2(T+\bar T)^{-x+y}|\phi_-|^2,
\end{equation*}
and 
\begin{equation*}
D_{\phi_+}W=\lambda e^{-aT}\phi_-,\qquad D_TW=D_{\phi_-}W=0. 
\end{equation*}
This F-term potential is re-expressed by using the normalized fields 
$\tilde\tau$ and $\tilde\phi_-$, and then, 
the complete form of the tree-level potential is given by 
\begin{equation*}
V_{\rm tree}=
\lambda^22^{\frac{-x+2y}2}\exp\left[-\sqrt2ae^{\sqrt{\frac2x}\tilde\tau}
+\sqrt{\frac2x}\left(-x+2y\right)\tilde\tau+|\tilde\phi_-|^2\right]
|\tilde\phi_-|^2
+\frac{g^2}2\left(\xi-|\tilde\phi_-|^2\right)^2. 
\end{equation*}
This potential was shown in Fig~\ref{fg:pot4} for the parameters given in Eq. (\ref{eq:in1}). 


Since we cannot find a SUSY vacuum in this potential 
with $\langle\phi_-\rangle\neq0$ (except for $\tilde\tau\rightarrow\pm\infty$), 
the CW potential  remains non-vanishing after the inflation. 
The inflaton direction, whose slope is dominated by the CW potential, 
has to be uplifted somehow. 

One simple way to solve this problem would be given 
by adding the following term to the superpotential, 
\begin{equation}
W\supset\lambda' e^{-b T}. \label{eq:add}
\end{equation}
We naively expect that 
the inflaton direction is uplifted without significant effects 
on the inflationary dynamics (for certain values of $\lambda'$ and $b$). 
We demonstrate this expectation for the case discussed in Section~3 (where the parameters are set as Eq.~(\ref{eq:in1}) and 
the potential is shown in Fig.~\ref{fg:pot4}). 
In the left panel of Fig.~\ref{fg:aft1}, 
we fix the additional parameters as 
$\lambda'/\sqrt\xi=0.3$ and $b=8\pi$ 
and draw the inflaton potential ($\tilde\phi_\pm=0$) 
with and without the above term (\ref{eq:add}) by using orange and blue lines, 
respectively. We see from this figure that 
the additional term can stabilize the inflaton direction without 
changing the inflationary dynamics. 
We also checked that the same result follows with 
another parameter set $\lambda'/\sqrt\xi=0.01$ and $b=-1$ 
as shown in the right panel (this parameter set corresponds to the positive exponent \footnote{Such a term may be generated by non-perturbative effects, such that non-perturbative effects depend on two or more moduli, but 
the moduli except $T$ are stabilized with heavy masses \cite{Abe:2005rx,Abe:2008xu}.}).

Alternatively, the inflaton direction may be stabilized without the additional term, 
because the F-component of $\phi_+$ cannot vanish and the SUSY breaking 
effect produces softmass terms for gaugino and sfermion fields. 
In that case, a new potential for the inflaton arises from 
quantum corrections only after the inflation so as to stabilize the inflaton. 
The details depend on specific models for the visible sector and 
we leave such a study for our future work.

\iffigure
\begin{figure}[t]
\centering
\includegraphics[width=8cm]{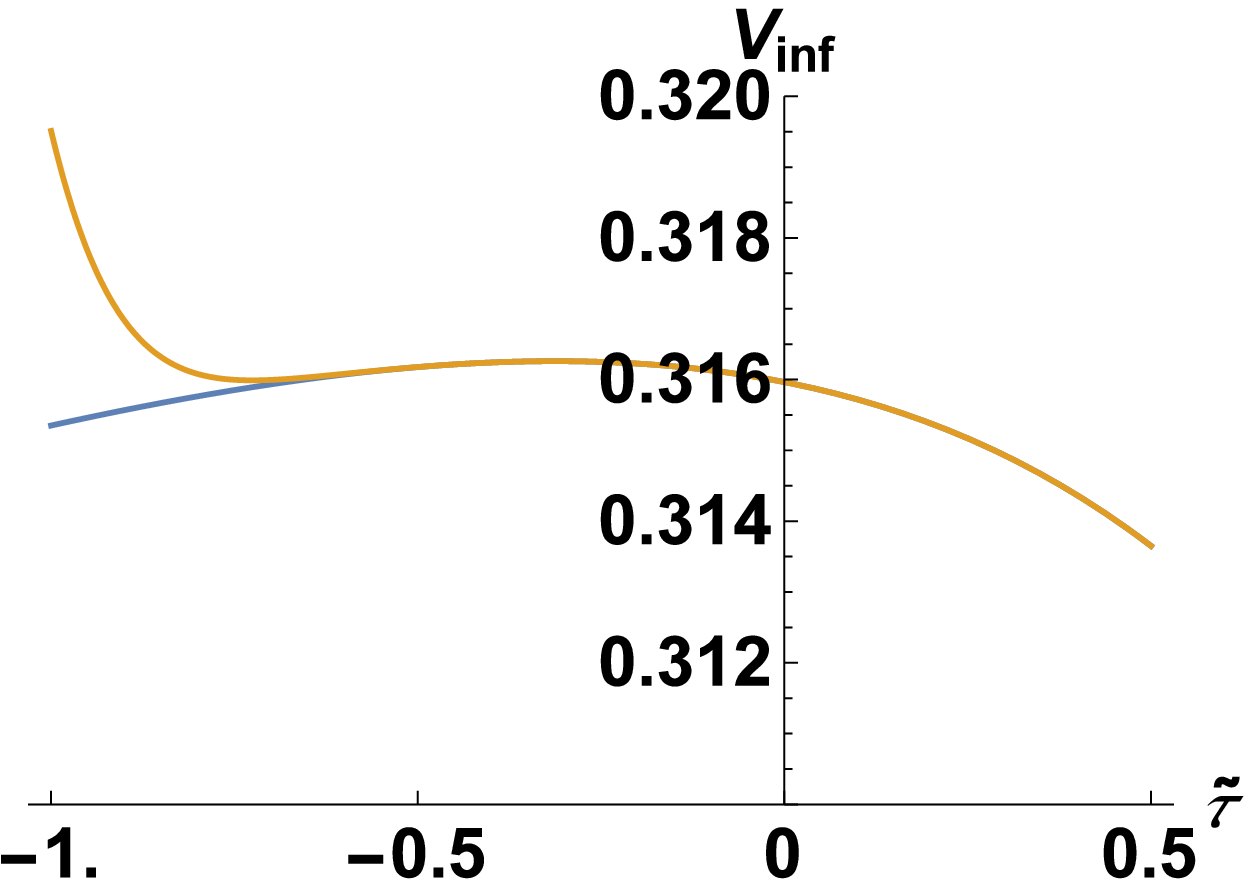}\hspace{0.5cm}
\includegraphics[width=8cm]{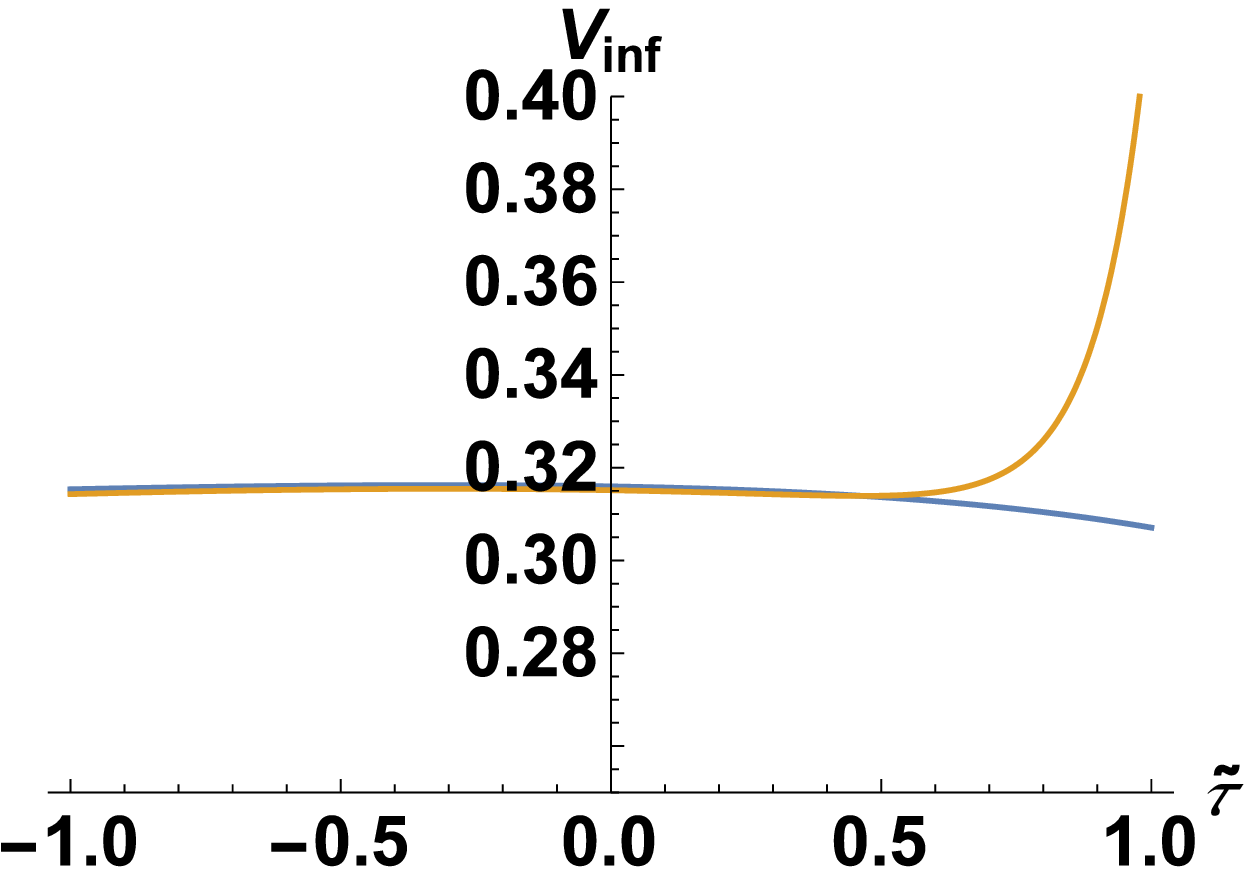}
\caption{
The normalized potential with (without) the additional term 
given by Eq.~(\ref{eq:add}) 
is shown by the orange (blue) line. 
The additional parameters 
are set as $(\lambda'/\sqrt\xi,b)=(0.3,8\pi)$ 
and $(\lambda'/\sqrt\xi,b)=(0.01,-1)$ in the left and right panels, 
respectively. 
}\label{fg:aft1}
\end{figure} 
\fi

\section{Conclusion and discussion}
We have proposed a new model for the D-term hybrid inflation. 
The CW potential possesses a flat enough direction around the top of a hill-like potential
which can lead to the desirable spectral tilt and suppress the cosmic string contribution to the CMB. 
We have found that our model is consistent with the latest CMB data for a wide range of
parameter space and, in particular, our model is consistent with the observations even when 
the parameters are set to natural and reasonable values of order unity. 
Notably, the $U(1)$ gauge coupling constant 
comparable to that of the Standard Model is possible in our scenario.

Our D-term inflation model can resolve the persistent inconsistencies among $A_s,n_s$ and the string tension in the conventional D-term inflation scenarios for the following reasons.
The crucial point is that our scenario can let $\epsilon$ and $\eta$ vary independently, 
thanks to the presence of the linear term in the inflaton potential.
This is in contrast to the conventional
D-term inflation where $\epsilon$ and $\eta$ are tightly related and their ratio is fixed for a given gauge coupling constant $|\epsilon/\eta|\approx  g^2/8 \pi^2$, so that
the string tension is fixed once the CMB power spectrum is given as shown in Eq.~(\ref{gmufixed}). 
 In our scenario, $A_s \sim V/(24 \pi^2 \epsilon)\sim g^2 \xi^2/(48 \pi^2 \epsilon) $ can be matched to the observed value by
adjusting $\epsilon$, while $n_s-1\sim 2 \eta$ can be tuned to the observed value by adjusting $\eta$ without changing $A_s$.
This lifting the tight relation between $\epsilon$ and $\eta$ stems from the characteristic form of the CW potential in
our scenario consisting of the linear and exponential terms. Our potential can enhance the second derivative of the potential while keeping
the first derivative small (so that $\eta\gg \epsilon$ even with an order unity gauge coupling), and the string tension cannot be fixed anymore simply by
fixing $n_s$ and $A_s$ as in the conventional D-term inflation model. 

The string tension predicted in our model is consistent with the Planck data.
However, such a string tension not too far from the current Planck bound in our scenario could potentially be probed with the future experiments such as the CMB and 21cm data achieving the sensitivity of order $G \mu \sim {\cal O}(10^{-8})$ \cite{dano2008,ciu2017,her2016,her2012,holder2010,kha2008,bern2010}.

There remains the CW potential even after the inflation, unless 
we design the additional term in the superpotential to induce a 
supersymmetric vacuum. 
The inflaton direction is unbounded because of the CW potential, 
which should be uplifted somehow. 
As discussed in the previous section, 
the uplifting is implemented by the additional terms to the superpotential. 
More interesting possibility is that the supersymmetry breaking is 
mediated to the visible sector and the mass splitting of 
the gauge superfields or matter superfields produces 
a one-loop potential for the inflaton. 
We will study these mechanisms in specific models in the future.

The inflation scale $H_{\rm inf}$ can be lowered by fine-tuning the initial condition 
and choosing a suppressed value of $\lambda$.
The reheating temperature consequently can be lower, whose precise nature, however, would depend on the uplifting mechanism discussed above and couplings to the visible sector.
We would study such aspects in concrete models elsewhere.

Finally, we comment on embedding our model into 
superstring theory.
Our K\"ahler potential can be realized by four-dimensional low-energy field theory 
derived from superstring theory, where $T$ would correspond to a modulus field.
Such a description would be valid for $\re T > 1$.
That is, the inflaton trajectory with 
$\tilde \tau >0$ would be interesting.
Also, the form of superpotential could be generated by non-perturbative effects such as 
D-brane instanton effects and gaugino condensation.
The typical value of $a$ would be $2 \pi$ when D-brane instanton generates such a superpotential.
On the other hand, a smaller value like $a \lesssim 1$ is required to lead to the successful inflation   
for $\tilde \tau >0$.
We need some mechanism to realize a smaller $a \lesssim 1$, e.g. by mixing moduli.
The realization of a non-vanishing FI term would be another important issue.
It may be generated by another modulus or induced dynamically by some strong interactions.
Thus, stringy realization of our D-term inflation model would be of great interest and deserve further study.

\section*{Acknowledgments}
We thank J. Evans and T. Sekiguchi for the useful discussions.
K.~K. was supported by IBS under the project code IBS-R018-D1.
T.~K. was is supported in part by the Grant-in-Aid for Scientific Research 
 No.~26247042  and No.~17H05395 from the Ministry of Education, Culture, Sports,
 Science and Technology in Japan.

\end{document}